\documentclass[a4paper,11pt]{article}
\usepackage{jheppub} 
\usepackage{lineno}

\usepackage{listings}
\usepackage{multirow}
\usepackage{makecell}
\usepackage{array}
\usepackage{subfigure}
\usepackage{siunitx} 
\usepackage{booktabs}

\usepackage{slashed}

\arxivnumber{xxxx.xxxx} 

\title{The LHC sensitivity to weak gauginos in light of the latest muon $g-2$ and dark matter results}

\author[a]{Shuang Liang,}
\author[b]{Kun Wang,}
\author[a]{Jun Zhao,}
\author[a]{and Jingya Zhu}
\affiliation[a]{School of Physics and Electronics, Henan University, Kaifeng 475004, China}
\affiliation[b]{College of Science, University of Shanghai for Science and Technology, Shanghai 200093,  China}
\emailAdd{liangshuang999@henu.edu.cn}
\emailAdd{kwang@usst.edu.cn}
\emailAdd{junzhao@henu.edu.cn}
\emailAdd{zhujy@henu.edu.cn}

\abstract{
Among the electroweakinos, the weak gauginos have the largest production rate at the LHC and should therefore be the primary focus of searches. In this work, we examine the LHC sensitivity to weak gauginos in light of the latest constraints from the muon $g-2$ anomaly and dark matter (DM) observations. To simultaneously account for the observed $5\sigma$ deviation in muon $g-2$ and the correct DM relic abundance, the DM candidate in the MSSM must be bino-like: wino- or higgsino-like neutralinos require TeV-scale masses to avoid underabundance, rendering the electroweakino spectrum too heavy to yield a sizable $g-2$ contribution. Moreover, tight direct detection limits disfavor light higgsinos, which can induce sizable DM–nucleon scattering via bino–higgsino mixing. We thus focus on scenarios with a bino-like LSP and relatively heavy higgsinos. 
Two classes of wino spectra are considered: a light-wino scenario (LWo) with $m_{\tilde{W}} \lesssim 1$~TeV, and a heavy-wino scenario (HWo) with $m_{\tilde{W}} \gtrsim 1$~TeV. For each, we delineate the viable parameter space under current constraints and assess the discovery potential at future LHC stages. We find that while the high-luminosity LHC (HL-LHC) can probe a portion of the parameter space, the high-energy 27~TeV LHC (HE-LHC) is capable of covering most of it.
}

\begin{document}
\maketitle
\flushbottom

\section{Introduction}
\label{sec:intro}

The longstanding discrepancy in the anomalous magnetic moment of the muon, $a_\mu \equiv (g-2)_\mu/2$, has been further solidified by the 2021 result of the Fermilab Muon $g-2$ Collaboration~\cite{Muong-2:2021ojo}, confirming the earlier E821 measurement at Brookhaven~\cite{Muong-2:2006rrc}. With the latest theoretical and experimental updates~\cite{Aoyama:2020ynm,RBC:2018dos,Muong-2:2023cdq, Muong-2:2024hpx}, the combined analysis now shows a $5.2\sigma$ deviation:
\begin{equation}
\Delta a_\mu = a_\mu^{\mathrm{exp}} - a_\mu^{\mathrm{SM}} = (2.49 \pm 0.48) \times 10^{-9},
\end{equation}
corresponding to a relative precision of 0.20 parts per million (ppm), and representing one of the most statistically significant hints of physics beyond the Standard Model (SM) to date.

In parallel, astrophysical and cosmological observations have firmly established the presence of non-luminous, non-baryonic dark matter (DM), accounting for about 27\% of the universe’s energy budget. Identifying its particle nature remains a fundamental open question in particle physics and cosmology. Weakly Interacting Massive Particles (WIMPs), thermally produced in the early universe, provide a compelling candidate class. These models naturally reproduce the observed relic density via the freeze-out mechanism, and have been extensively tested through direct, indirect, and collider-based searches~\cite{Arcadi:2017kky,Chakraborti:2022sbj,Chakraborti:2021kkr,Cox:2018qyi,Tang:2022pxh,Cao:2018rix,Cao:2022ovk,Barman:2022jdg,Wang:2020xta,Wang:2020dtb,Li:2023kbf,Wang:2024ozr,Chakraborti:2021bmv}.

Reconciling the muon $g–2$ anomaly with the observed dark matter relic abundance and the null results from LHC supersymmetry searches presents a complex and highly constrained challenge for the minimal supersymmetric standard model (MSSM)~\cite{Haber:1984rc,Martin:1997ns,Djouadi:2005gj,Cao:2011sn,Wang:2018vrr,Baer:2020kwz,Yang:2022qyz,Wang:2022rfd,Dong:2024juh}. To accommodate the $g$–2 discrepancy, the spectrum must feature relatively light electroweakinos and sleptons, especially charginos and neutralinos with wino or higgsino components, and left-handed smuons or muon sneutrinos~\cite{Moroi:1995yh,Martin:2001st,Stockinger:2006zn,Athron:2015rva,Chakraborti:2020vjp,Chakraborti:2021dli,Chakraborti:2021squ,Chakraborti:2021ynm,Chakraborti:2021mbr,Athron:2021iuf,Wang:2021bcx,Ning:2017dng,Abdughani:2021pdc,Cao:2021tuh,Wang:2021lwi,Cao:2022htd,Cao:2022chy,Zhao:2021eaa,Yang:2021duj,Zhang:2021gun,Wang:2018vxp,Li:2023tlk,Cao:2019evo,Cao:2021lmj,Wang:2022wdy,Li:2021poy,Aboubrahim:2021phn}. These states generate sizable loop contributions to $\Delta a_\mu$, potentially accounting for the observed deviation.
On the other hand, satisfying the relic density and direct detection bounds typically requires a bino-like neutralino as the lightest supersymmetric particle (LSP), with minimal higgsino or wino admixture. A nearly pure bino, however, annihilates too inefficiently in the early universe, often resulting in overabundant dark matter unless supplemented by efficient depletion channels such as coannihilation. Achieving such coannihilation typically demands compressed spectra and fine-tuned mass splittings, especially when involving winos or sleptons.
In addition, current LHC searches place stringent lower bounds on the masses of charginos and sleptons, particularly in the presence of light bino-like LSPs with sizable production cross sections. These competing requirements introduce significant tension into the viable MSSM parameter space and motivate a detailed investigation of the interplay between them.

In this work, we investigate the LHC sensitivity to weak gauginos in the MSSM, taking into account the latest constraints from the muon $g–2$ anomaly and dark matter observations. 
We focus on two well-motivated classes of scenarios that can potentially satisfy all current experimental bounds: one in which a bino-like neutralino coannihilates with a slightly heavier wino, and another involving coannihilation with nearly mass-degenerate sleptons.
These configurations are systematically explored through dedicated scans of the MSSM parameter space, identifying viable regions that reconcile the muon $g–2$ discrepancy, the observed dark matter relic abundance, and null results from direct detection and collider experiments.
Although related scenarios have been studied previously~\cite{Cho:2011rk, Endo:2013bba, Abdughani:2019wai, Ajaib:2015yma, Badziak:2014kea, Sabatta:2019nfg, Zhao:2022pnv}, many analyses either neglected the requirement of thermal relic matching, focused only on current LHC bounds, or did not consider the potential reach of future LHC upgrades. We go beyond previous work by evaluating the prospects for probing these scenarios at both the High Luminosity (HL-) and High Energy (HE-) phases of the LHC, highlighting characteristic spectra and providing benchmark points to inform experimental search strategies.

The structure of this paper is as follows.  
In Sec.~\ref{sec:scenario}, we briefly review the status of the muon $g$–2 anomaly and dark matter in the context of the MSSM, highlighting the relevant mechanisms that can simultaneously address both phenomena.  
In Sec.~\ref{sec:scan}, we describe the parameter scan strategy and discuss the experimental constraints used to identify viable MSSM scenarios.  
In Sec.~\ref{sec:LHC}, we analyze the LHC sensitivity to weak gauginos in the HL-LHC and HE-LHC phases, focusing on the discovery potential of the benchmark scenarios.  
Finally, we summarize our findings in Sec.~\ref{sec:conclusions}.

\section{\label{sec:scenario}Overview of the MSSM framework}

In the MSSM, there are four neutralinos, denoted by $\tilde{\chi}^0_{1,2,3,4}$ in ascending order of mass. These states are formed from linear combinations of the bino ($\tilde{B}$), wino ($\tilde{W}^0$), and the neutral higgsinos ($\tilde{H}_u^0$, $\tilde{H}_d^0$). Their masses and mixings are governed by the neutralino mass matrix, which can be written as~\cite{Haber:1984rc}:
\begin{eqnarray}
	M_{\tilde{\chi}^0} &=& \left(\begin{array}{cccc}
		M_1 & 0 & -c_{\beta} s_W m_Z & s_{\beta} s_W  m_Z \\
		0 & M_2 & c_{\beta} c_W m_Z & -s_{\beta} c_W m_Z  \\
		-c_{\beta} s_W m_Z & c_{\beta} c_W m_Z & 0 & -\mu  \\
		s_{\beta} s_W m_Z & -s_{\beta} c_W m_Z& -\mu & 0
	\end{array} \right)
	\label{neutralinomassmatrix}
\end{eqnarray}

\noindent Here, $s_\beta$, $c_\beta$, $s_W$, and $c_W$ represent $\sin\beta$, $\cos\beta$, $\sin\theta_W$, and $\cos\theta_W$, respectively.
The parameters $M_1$ and $M_2$ denote the soft SUSY-breaking mass terms for the bino and wino, respectively, while $\mu$ is the higgsino mass parameter.
The mass matrix is diagonalized by a unitary $4 \times 4$ matrix $N$.

Similarly, the chargino mass matrix, arising from the mixing of the charged wino ($\tilde{W}^\pm$) and the charged higgsinos ($\tilde{H}_d^-$, $\tilde{H}_u^+$), is given by:
\begin{eqnarray}
		M_{\tilde{\chi}^\pm}&=&\left( \begin{array}{cc}
				M_2& \sqrt{2} s_{\beta} m_W \\
				\sqrt{2} c_{\beta} m_W& \mu
			\end{array} \right)
		\label{mass}
\end{eqnarray}
This matrix can be diagonalized by two unitary $2 \times 2$ matrices, $U$ and $V$.

\begin{figure*}[htbp] 
    \centering 
    \includegraphics[width=.7\linewidth]{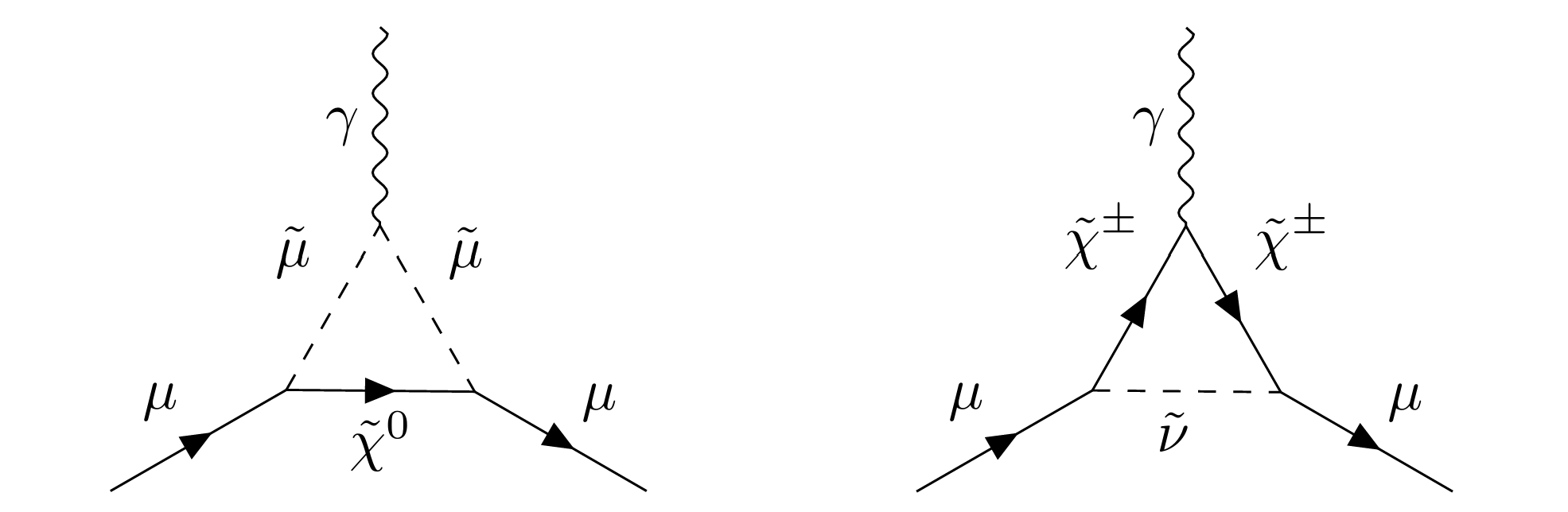}\vspace{ 0.0cm} 
    \caption{One-loop diagrams contributing to the SUSY corrections to $a_\mu$.
} 
\label{fig:loop}
\end{figure*}

The dominant supersymmetric contributions to the muon $g-2$ originate from one-loop diagrams involving neutralino–smuon ($\tilde{\chi}^0$–$\tilde{\mu}$) and chargino–sneutrino ($\tilde{\chi}^\pm$–$\tilde{\nu}$) exchanges, as illustrated in Fig.~\ref{fig:loop}. At the one-loop level, the resulting MSSM contributions to $a_\mu$ can be written as \cite{Martin:2001st}:
\begin{equation}
    \delta a_{\mu}^{\text{SUSY}} = \delta a_{\mu}^{\tilde{\chi}^0 \tilde{\mu}} + \delta a_{\mu}^{\tilde{\chi}^\pm \tilde{\nu}}\, ,
\end{equation}
with
\begin{align}
	\delta a_\mu^{\tilde{\chi}^0\tilde{\mu}} =& \frac{m_\mu}{16\pi^2}
	\sum_{i,m}\left\{ -\frac{m_\mu}{ 12 m^2_{\tilde\mu_m}}
	(|n^L_{im}|^2+ |n^R_{im}|^2)F^N_1(x_{im})
	+\frac{m_{\tilde{\chi}^0_i}}{3 m^2_{\tilde \mu_m}}
	{\rm Re}[n^L_{im}n^R_{im}] F^N_2(x_{im})\right\} \nonumber\\
	=& \frac{m_\mu}{16\pi^2} \sum_{i,m} \left\{ -\frac{m_\mu\left( |\eta_{im}^L|^2 + |\eta_{im}^R|^2 \right)}{6m_{\tilde{\mu}_m}^2(1-x_{im})^4}    \left[ 1 - 6x_{im} + 3x_{im}^2 + 2x_{im}^3 - 6x_{im}^2 \ln x_{im} \right] \right. \nonumber \\
	& \left. + \frac{m_{\tilde{\chi}_i^0}}{3m_{\tilde{\mu}_m}^2} \operatorname{Re}[\eta_{im}^L \eta_{im}^R]  \frac{3}{(1-x_{im})^3} \left[ 1 - x_{im}^2 + 2x_{im} \ln x_{im} \right] \right\} \, , 
	\\[2ex] 
	\delta a_\mu^{\tilde{\chi}^\pm\tilde{\mu}} =& \frac{m_\mu}{16\pi^2}\sum_k
	\left\{ \frac{m_\mu}{ 12 m^2_{\tilde\nu_\mu}}
	(|c^L_k|^2+ |c^R_k|^2)F^C_1(x_k)
	+\frac{2m_{\tilde{\chi}^\pm_k}}{3m^2_{\tilde\nu_\mu}}
	{\rm Re}[ c^L_kc^R_k] F^C_2(x_k)\right\}\nonumber\\
	=& \frac{m_\mu}{16\pi^2} \sum_k \left\{ \frac{m_\mu\left( |c_k^L|^2 + |c_k^R|^2 \right)}{6m_{\tilde{\nu}_\mu}^2(1-x_k)^4}   \left[ 2 + 3x_k - 6x_k^2 + x_k^3 + 6x_k \ln x_k \right] \right. \nonumber \\
	& \left. + \frac{2m_{\chi_k^\pm}}{3m_{\tilde{\nu}_\mu}^2} \operatorname{Re}[c_k^L c_k^R]  \left( -\frac{3}{2(1-x_k)^3} \right) \left[ 3 - 4x_k + x_k^2 + 2 \ln x_k \right] \right\} \, , 
\end{align}

\noindent where $i=1,2,3,4$, $m=1,2$, and $k=1,2$ label the mass eigenstates of neutralinos, smuons, and charginos, respectively. The loop functions $F^N_{1,2}$ and $F^C_{1,2}$ depend on the mass ratios $x_{im} = m^2_{\tilde{\chi}^0_i} / m^2_{\tilde{\mu}m}$ and $x_k = m^2{\tilde{\chi}^\pm_k} / m^2_{\tilde{\nu}\mu}$, and are normalized such that $F^N{1,2}(1) = F^C_{1,2}(1) = 1$. The couplings $n^R_{im}$, $n^L_{im}$, $c^R_k$, and $c^L_k$ are defined as:
\begin{eqnarray}
&& n^R_{im}  =   \sqrt{2} g_1 N_{i1} X_{m2} + y_\mu N_{i3} X_{m1} ,\\
&& n^L_{im}  =   {1\over \sqrt{2}} \left (g_2 N_{i2} + g_1 N_{i1}
               \right ) X_{m1}^* - y_\mu N_{i3} X^*_{m2} ,\\
&& c^R_k  =  y_\mu U_{k2} ,~~~~ c^L_k  =  -g_2V_{k1} ,
\end{eqnarray}
where $y_\mu = \dfrac{g_2 m_\mu}{\sqrt{2} m_W \cos\beta}$ denotes the muon Yukawa coupling.

Assuming a common mass scale $M_{\rm SUSY}$ for all supersymmetric particles, the contributions from the $\tilde{\chi}^0$–$\tilde{\mu}$ and $\tilde{\chi}^\pm$–$\tilde{\nu}$ loops can be approximated as follows \cite{Moroi:1995yh}:
\begin{align}
\delta a_\mu^{\tilde{\chi}^0\tilde{\mu}} &= \frac{1}{192\pi^2} \frac{m_\mu^2}{M_{\text{SUSY}}^2} (g_1^2 - g_2^2) \tan \beta\, ,\\
\delta a_\mu^{\tilde{\chi}^\pm\tilde{\mu}} &= \frac{1}{32\pi^2} \frac{m_\mu^2}{M_{\text{SUSY}}^2} g_2^2 \tan \beta\, ,
\end{align}
and the corresponding SUSY contributions to $a_\mu$ are approximately given by:
\begin{eqnarray}
\delta a_\mu^{\rm SUSY} 
= 14 \tan\beta \left ( \frac{100~ {\rm GeV}}{M_{\rm SUSY}} \right )^2 10^{-10} .
\label{msusy}
\end{eqnarray}
From the above expression, it is evident that the dominant contribution arises from the $\tilde{\chi}^\pm$–$\tilde{\nu}$ loops. The SUSY contribution is enhanced by a larger value of $\tan\beta$, whereas it is suppressed by the overall mass scale of the superpartners. Therefore, to achieve a sizable effect on the muon $g–2$, the masses of the relevant charginos, neutralinos, and sleptons must remain sufficiently light.

In this work, we require the first neutralino $\tilde{\chi}_1^0$ to be the lightest SUSY particle (LSP), which serves as the DM candidate. 
We assume the lightest neutralino $\tilde{\chi}_1^0$ to be the lightest supersymmetric particle (LSP), thereby serving as the dark matter candidate. When $\tilde{\chi}_1^0$ is wino-like or higgsino-like, its (co-)annihilation rate in the early universe is typically too efficient to yield the observed relic abundance. To match the correct dark matter density without invoking additional non-SUSY components such as axions, such candidates would need to have masses at the TeV scale~\cite{Arkani-Hamed:2006wnf, Han:2013usa}. However, this scenario results in an electroweakino spectrum that is too heavy to account for the observed muon $g–2$ anomaly. Furthermore, wino- or higgsino-like neutralinos generally exhibit sizable scattering cross-sections with nucleons, making them strongly constrained by current direct detection limits.

\section{\label{sec:scan}Constraining the MSSM with muon $g–2$ and dark matter data}

We investigate two well-motivated scenarios within the MSSM: the light-wino scenario (LWo) and the heavy-wino scenario (HWo), which are distinguished by the mass hierarchy between the wino and bino. To explore the viable parameter space consistent with current experimental results, we perform comprehensive numerical scans using a suite of established computational tools.

We generate the SUSY mass spectrum using the \texttt{SUSY-HIT} package~\cite{Djouadi:2006bz}, with the spectrum calculation handled by the \texttt{SuSpect-2.41} subroutine~\cite{Djouadi:2002ze}. Higgs boson decays and SUSY particle decays are computed with \texttt{HDECAY-3.4}~\cite{Djouadi:1997yw, Djouadi:2018xqq} and \texttt{SDECAY-1.5}~\cite{Muhlleitner:2003vg}, respectively. The dark matter relic density of the lightest neutralino is evaluated using \texttt{MicrOMEGAs-5.3.41}~\cite{Belanger:2010gh}, while the two-loop SUSY contributions to the muon anomalous magnetic moment are computed via \texttt{GM2Calc-2.2.0}~\cite{Athron:2015rva, Athron:2022gga}. 
Finally, collider constraints from existing LHC searches are implemented using the \texttt{SModelS-3.0.0} framework~\cite{Altakach:2024jwk, Kraml:2013mwa, Ambrogi:2017neo, Ambrogi:2018ujg, Alguero:2021dig}, which decomposes the SUSY spectrum into simplified topologies and compares them against the latest experimental bounds.

The parameter ranges used in our scan are defined as follows. For the light-wino scenario (LWo), we take
\begin{align}
0~{\rm TeV} \leq M_1 \simeq M_2 \leq 1~{\rm TeV}, \quad 
3~{\rm TeV} \leq \mu \leq 5~{\rm TeV},
\end{align}
while for the heavy-wino scenario (HWo), we consider
\begin{align}
0~{\rm TeV} \leq M_1 \leq 1~{\rm TeV}, \quad
1~{\rm TeV} \leq M_2 \leq 5~{\rm TeV}, \quad
3~{\rm TeV} \leq \mu \leq 5~{\rm TeV}.
\end{align}
Here, $M_1$, $M_2$, and $\mu$ denote the soft mass parameters of the bino, wino, and higgsino, respectively. 
For the HWo scenario, we further impose $M_2 \leq \mu$ to ensure that the second-lightest neutralino $\tilde{\chi}_2^0$ is predominantly wino-like.

In both scenarios, the remaining input parameters are scanned over the following ranges:
\begin{align}
200~{\rm GeV} \leq M_{L_\ell} = M_{E_\ell} \leq 800~{\rm GeV}, \quad \\
3~{\rm TeV} \leq M_{\tilde{t}_R} \leq 5~{\rm TeV}, \quad
1 \leq \tan\beta \leq 50, \\
-5~{\rm TeV} \leq A_{t,b,\tau} \leq 5~{\rm TeV}, \quad
A_{u,d,e} = 0,
\end{align}
with all other soft mass parameters fixed at 5 TeV.

In our scan, we consider the following experimental constraints:
\begin{itemize}
\item[(1)] 
The mass of the SM-like Higgs boson is required to lie within the range $122~\mathrm{GeV} < m_h < 128~\mathrm{GeV}$, accounting for theoretical uncertainties associated with two-loop corrections evaluated under the approximations of vanishing external momenta and negligible electroweak gauge couplings. Additionally, the Higgs signal strengths must be consistent with the results from LHC measurements, as implemented in \texttt{HiggsBounds-5.10.0}~\cite{Bechtle:2020pkv, Bechtle:2013wla, Bechtle:2011sb, Bechtle:2008jh, Bahl:2022igd} and \texttt{HiggsSignals-2.6.2}~\cite{Bechtle:2013xfa, Bechtle:2020uwn, Bahl:2022igd}.

\item[(2)] 
We impose the metastability condition for the electroweak vacuum, which requires
\begin{align}
    |A_t| \lesssim 2.67 \sqrt{M_{\tilde{Q}_{3L}}^2 + M_{\tilde{t}_R}^2 + M_A^2 \cos^2 \beta},
\end{align}
to prevent dangerous charge- and color-breaking minima~\cite{Chowdhury:2013dka}.

\item[(3)] 
The masses of sleptons are required to exceed 200~GeV, consistent with exclusion limits from LEP2 and LHC searches.

\item[(4)] 
The dark matter relic density must lie within the $2\sigma$ range of the Planck 2018 result, $\Omega_{\mathrm{DM}} h^2 = 0.120 \pm 0.001$~\cite{Planck:2018vyg}.

\item[(5)] 
The SUSY contributions to the anomalous magnetic moment of the muon must account for the measured deviation from the SM prediction, 
\begin{align}
\Delta a_{\mu} \equiv a_{\mu}^{\mathrm{exp}} - a_{\mu}^{\mathrm{SM}} = (2.49 \pm 0.48) \times 10^{-9},
\end{align}
within a $2\sigma$ confidence level~\cite{Muong-2:2024hpx}.

\item[(6)] 
We impose constraints from current dark matter direct detection experiments, including bounds on the spin-independent cross section from XENONnT and related experiments~\cite{XENON:2023cxc,PandaX:2024qfu,LZCollaboration:2024lux,PandaX:2022xas,LZ:2018qzl,XENON:2020kmp,PANDA-X:2024dlo}.

\item[(7)] 
Collider constraints from supersymmetry searches at the LHC are incorporated via the \texttt{SModelS-3.00} package~\cite{Altakach:2024jwk, Kraml:2013mwa, Ambrogi:2017neo, Ambrogi:2018ujg, Alguero:2021dig}.
\end{itemize}

\begin{figure*}[htbp] 
    \centering 
    \includegraphics[width=.9\linewidth]{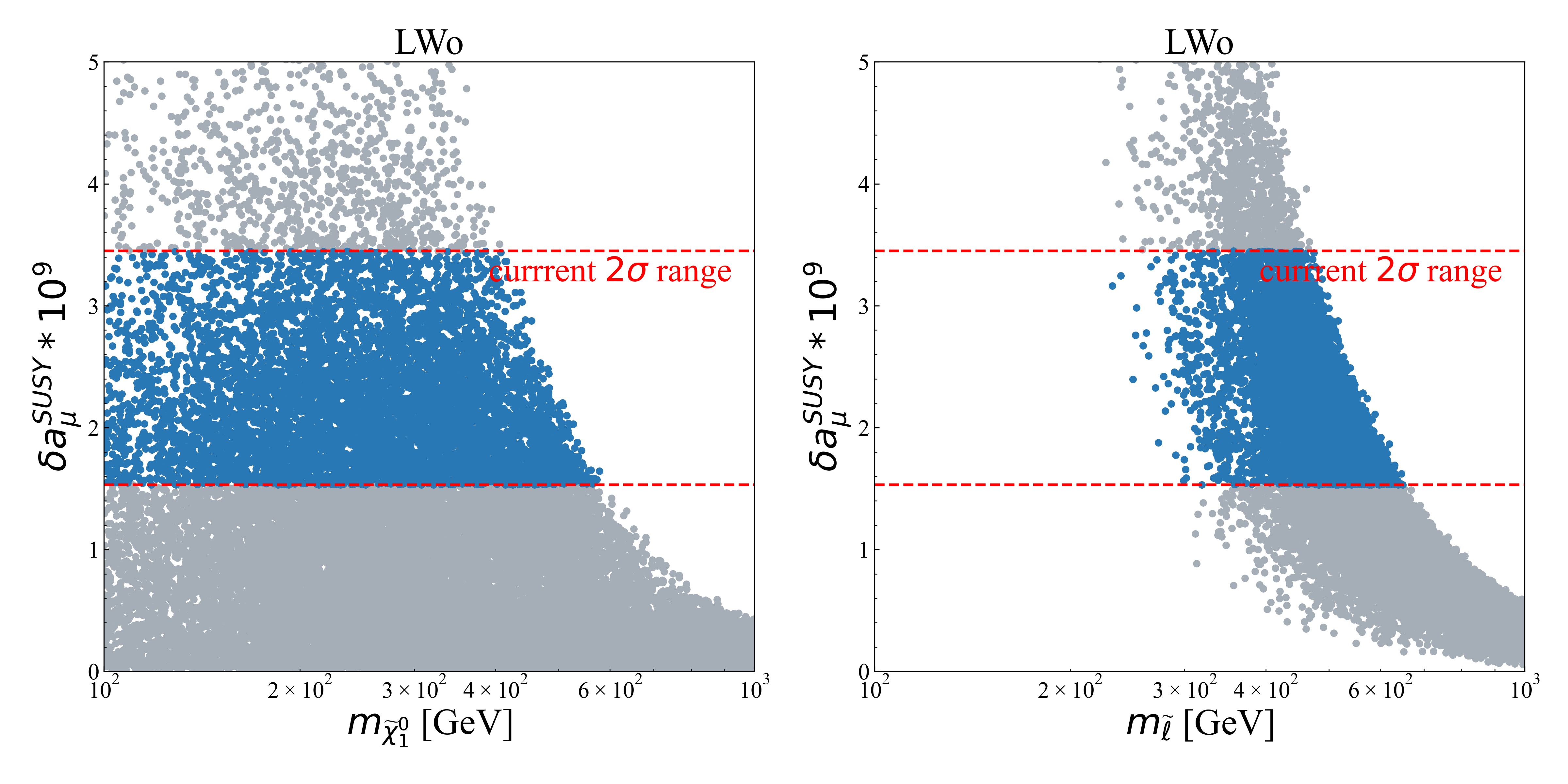}\hspace{.14\linewidth}
    \includegraphics[width=.9\linewidth]{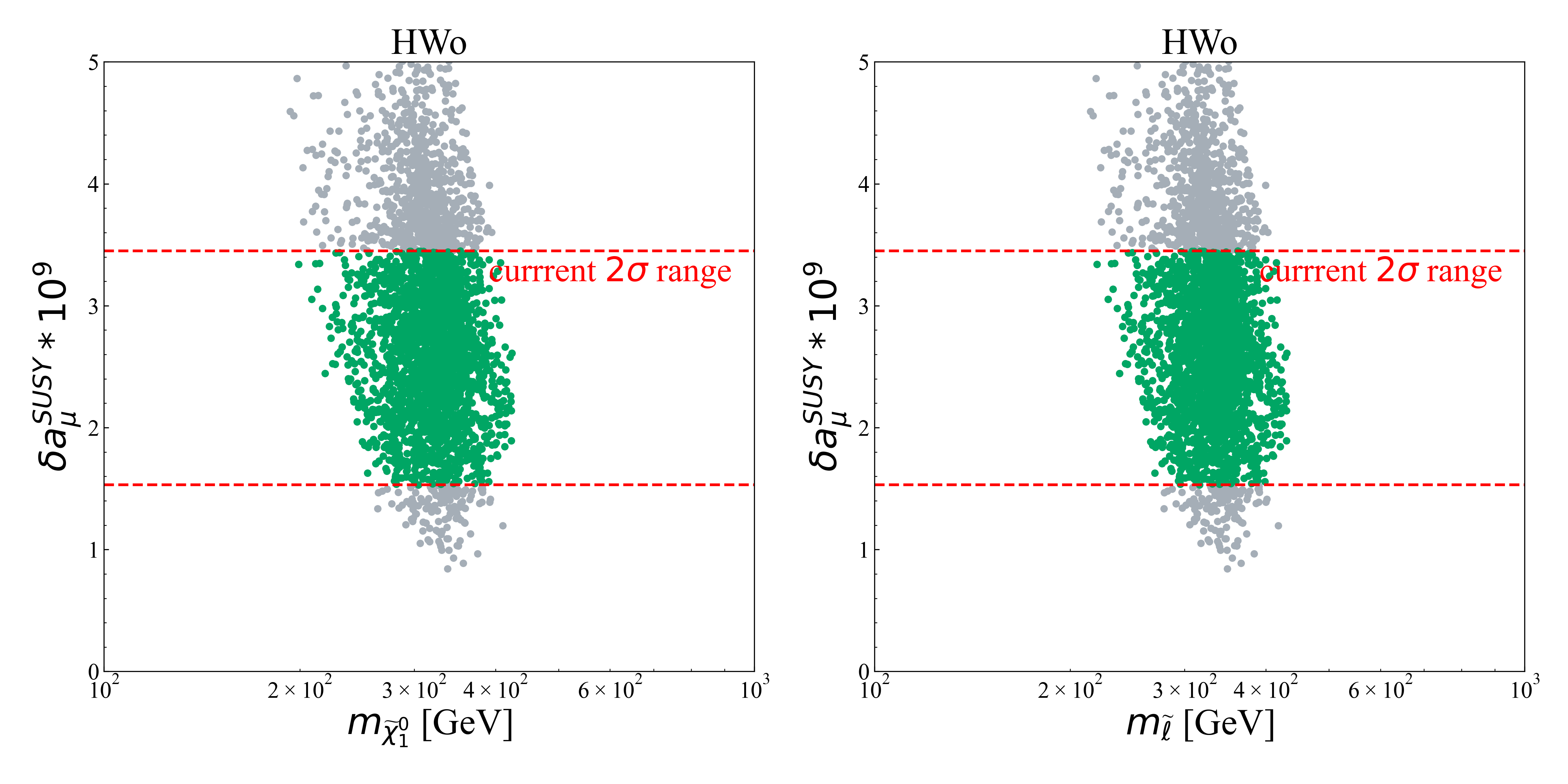}
\vspace{ 0.0cm} 
\caption{
Scatter plots of the two sample categories satisfying constraints (1)–(4), shown in the planes of $\delta a_\mu^{\mathrm{SUSY}}$ versus $m_{\tilde{\chi}_1^0}$ and $m_{\tilde{\ell}}$, respectively. The regions enclosed by dotted lines indicate the current $2\sigma$ range for the muon anomalous magnetic moment.
}
\label{fig:mass}
\end{figure*}

In Fig.~\ref{fig:mass}, we present scatter plots of the samples that satisfy constraints (1)–(4), projected onto the $\delta a_\mu^{\mathrm{SUSY}}$ versus $m_{\tilde{\chi}_1^0}$ and $m_{\tilde{\ell}}$ planes. The blue and green regions correspond to the $2\sigma$ ranges within which the LWo and HWo scenarios, respectively, can account for the muon $g$–2 anomaly. The key features are summarized as follows:
\begin{itemize}
  \item  \textbf{For the LWo scenario}:
  \begin{itemize}
    \item The bino-like $\tilde{\chi}_1^0$ and the sleptons must be lighter than approximately 580~GeV and 650~GeV, respectively, to accommodate the muon $g$–2 discrepancy.
    \item The observed relic abundance is primarily realized through coannihilation involving the bino-like $\tilde{\chi}_1^0$, the wino-like $\tilde{\chi}_2^0$, and the chargino $\tilde{\chi}_1^\pm$.
  \end{itemize}

  \item \textbf{For the HWo scenario}:
  \begin{itemize}
    \item Both $\tilde{\chi}_1^0$ and the sleptons are required to be lighter than approximately 430~GeV in order to explain the muon $g$–2 anomaly.
    \item The correct relic density is achieved via coannihilation between the bino-like LSP and nearly degenerate sleptons.
  \end{itemize}
\end{itemize}

\begin{figure*}[htbp] 
    \centering 
    \includegraphics[width=.9\linewidth]{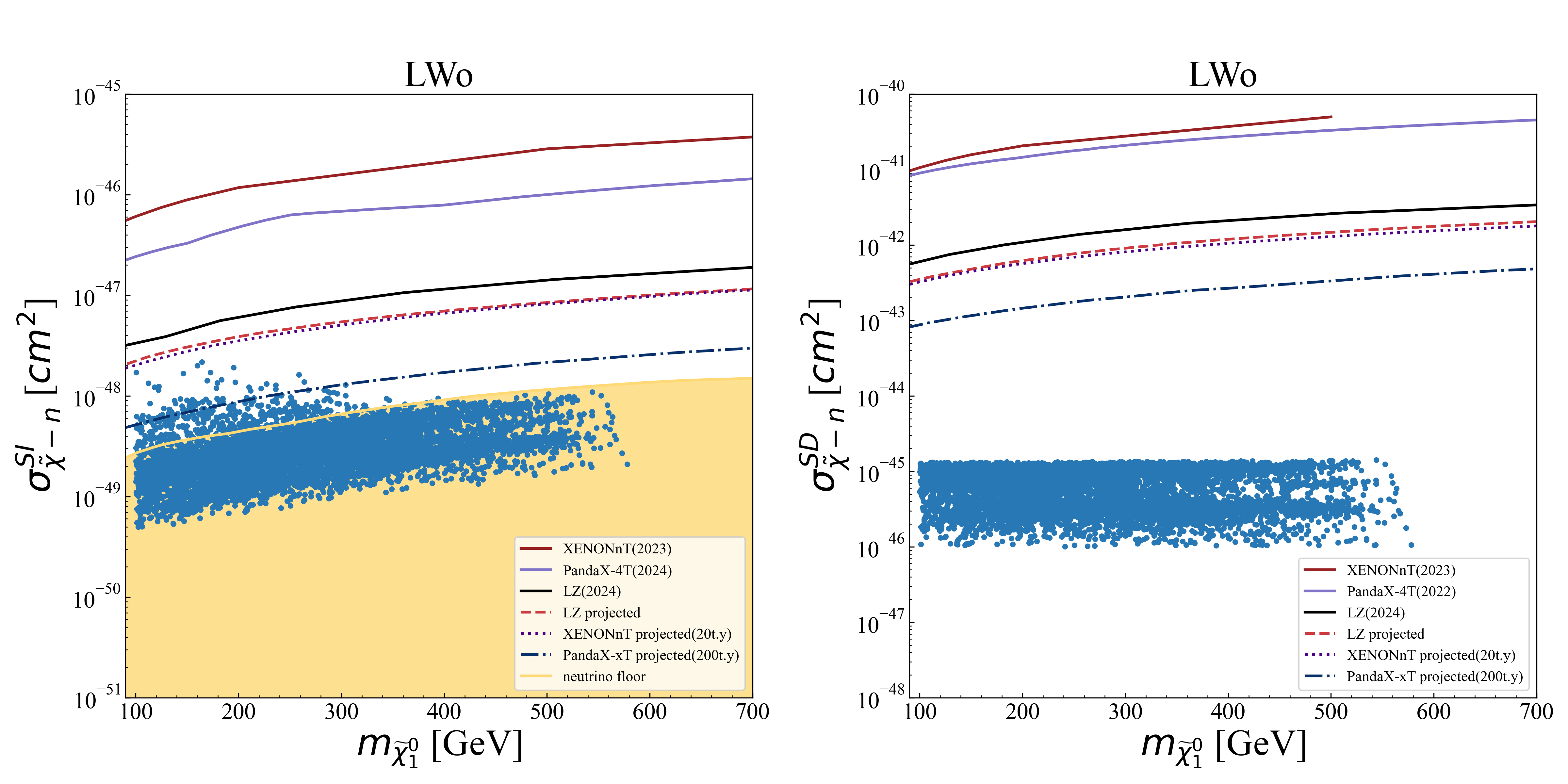}\hspace{.07\linewidth}
    \includegraphics[width=.9\linewidth]{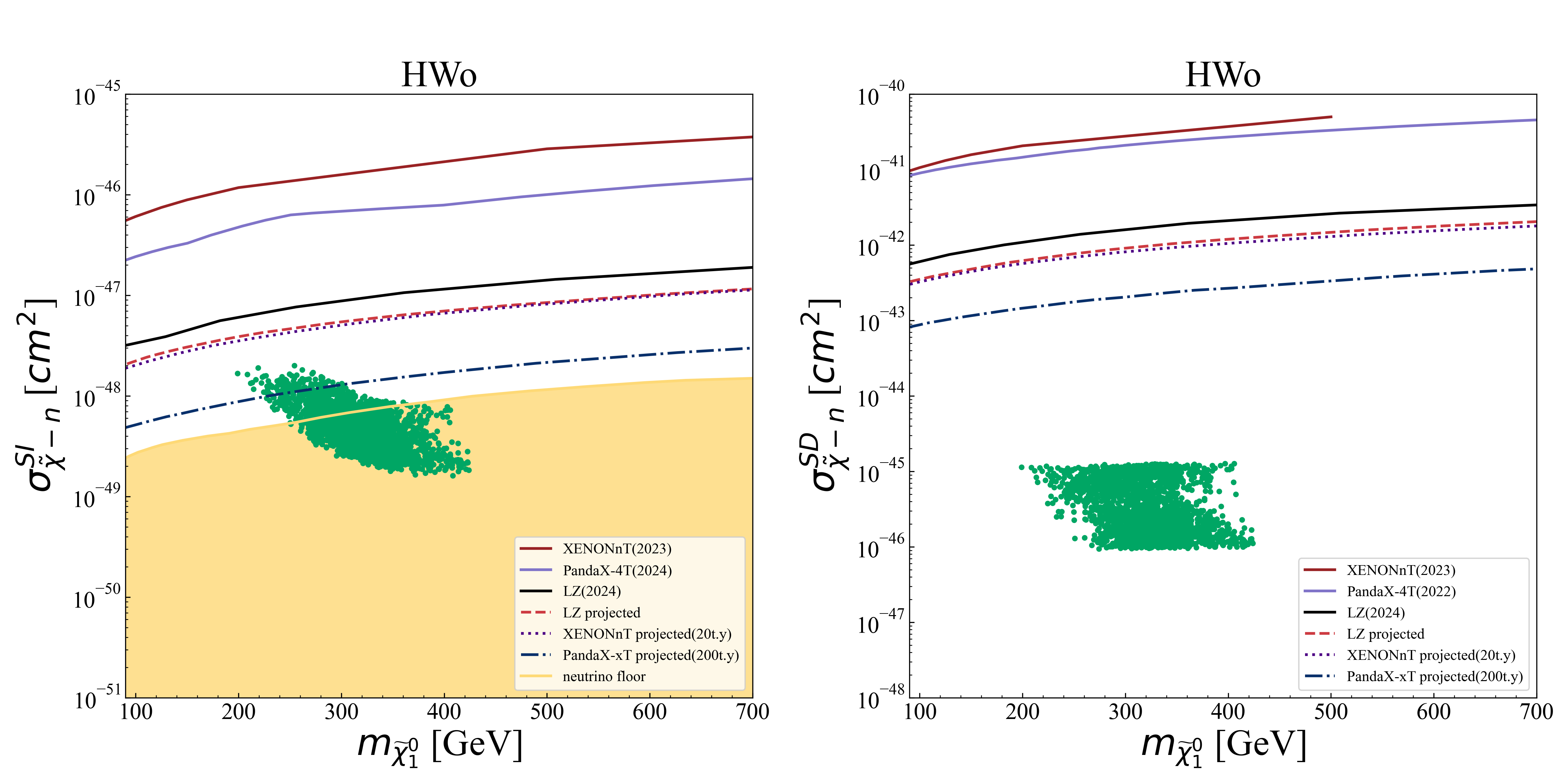}
\vspace{ 0.0cm} 
\caption{
Scatter plots of the samples that satisfy constraints (1)–(5), showing the spin-independent and spin-dependent scattering cross sections of the lightest neutralino (LSP) with nucleons as a function of the LSP mass.
The $90\%$ C.L. upper limits from recent direct detection experiments, including XENONnT-2023~\cite{XENON:2023cxc}, PandaX-4T-2024~\cite{PandaX:2024qfu}, LZ-2024~\cite{LZCollaboration:2024lux}, and PandaX-4T-2022~\cite{PandaX:2022xas}, are overlaid.
Projected sensitivities from future experiments, LZ~\cite{LZ:2018qzl}, XENONnT (20 ton-year)\cite{XENON:2020kmp}, and PandaX-xT (200 ton-year)\cite{PANDA-X:2024dlo}, are also shown.
The orange shaded region represents the neutrino floor~\cite{Billard:2013qya}, below which coherent neutrino scattering becomes a limiting background.
}
\label{fig:SISD}
\end{figure*}

In Fig.~\ref{fig:SISD}, we present the spin-independent (SI) and spin-dependent (SD) scattering cross sections of the lightest neutralino (LSP) with nucleons for the samples that satisfy constraints (1)–(5). In both scenarios, the LSP is predominantly bino-like, leading to highly suppressed couplings to quarks and consequently small SI and SD cross sections. These are well below the sensitivity of current direct detection experiments. Only a small fraction of the parameter space is expected to be probed by the next generation of dark matter experiments.

\begin{figure*}[htbp] 
    \centering 
    \includegraphics[width=.9\linewidth]{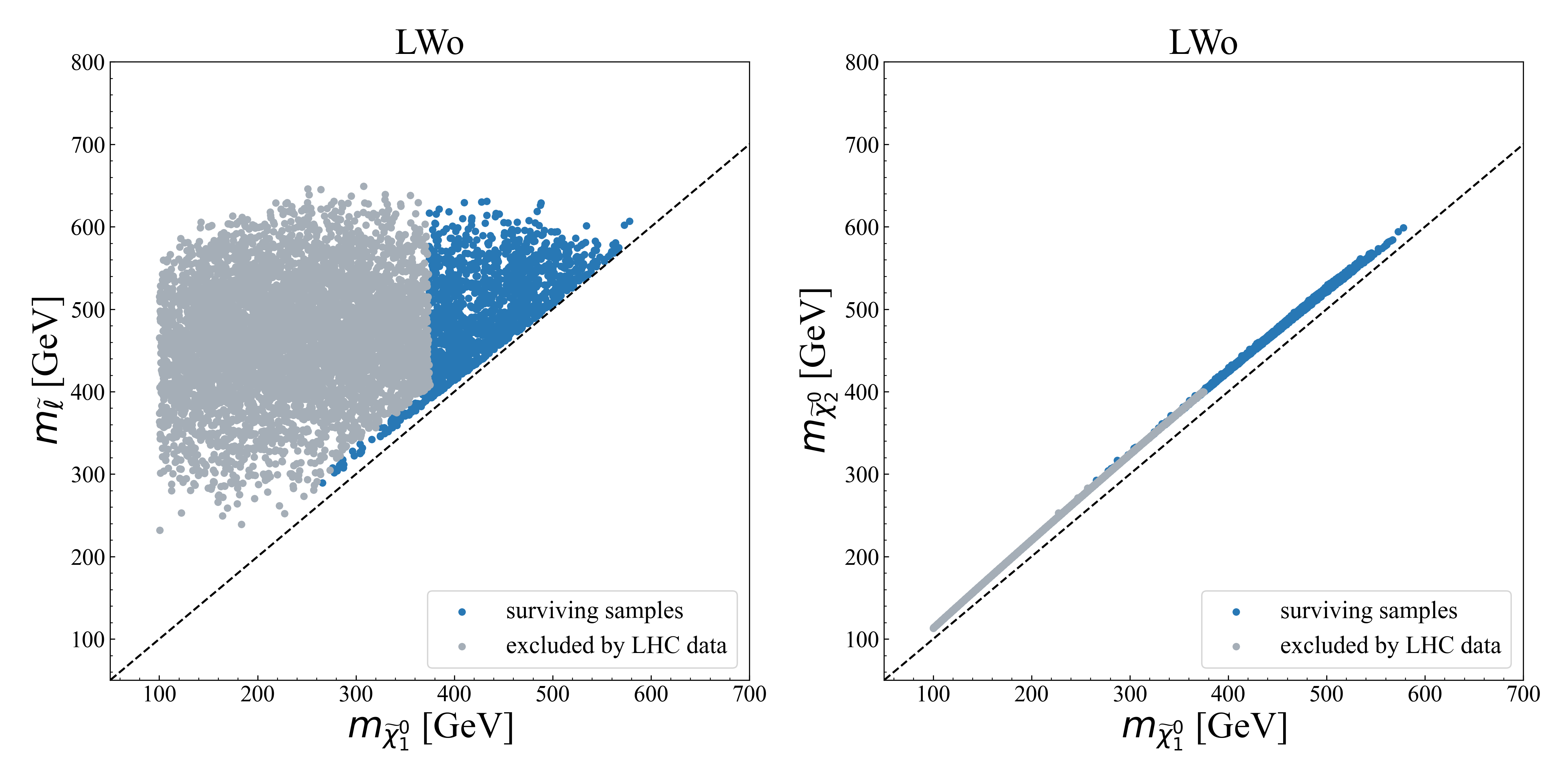}\hspace{.14\linewidth}
    \includegraphics[width=.9\linewidth]{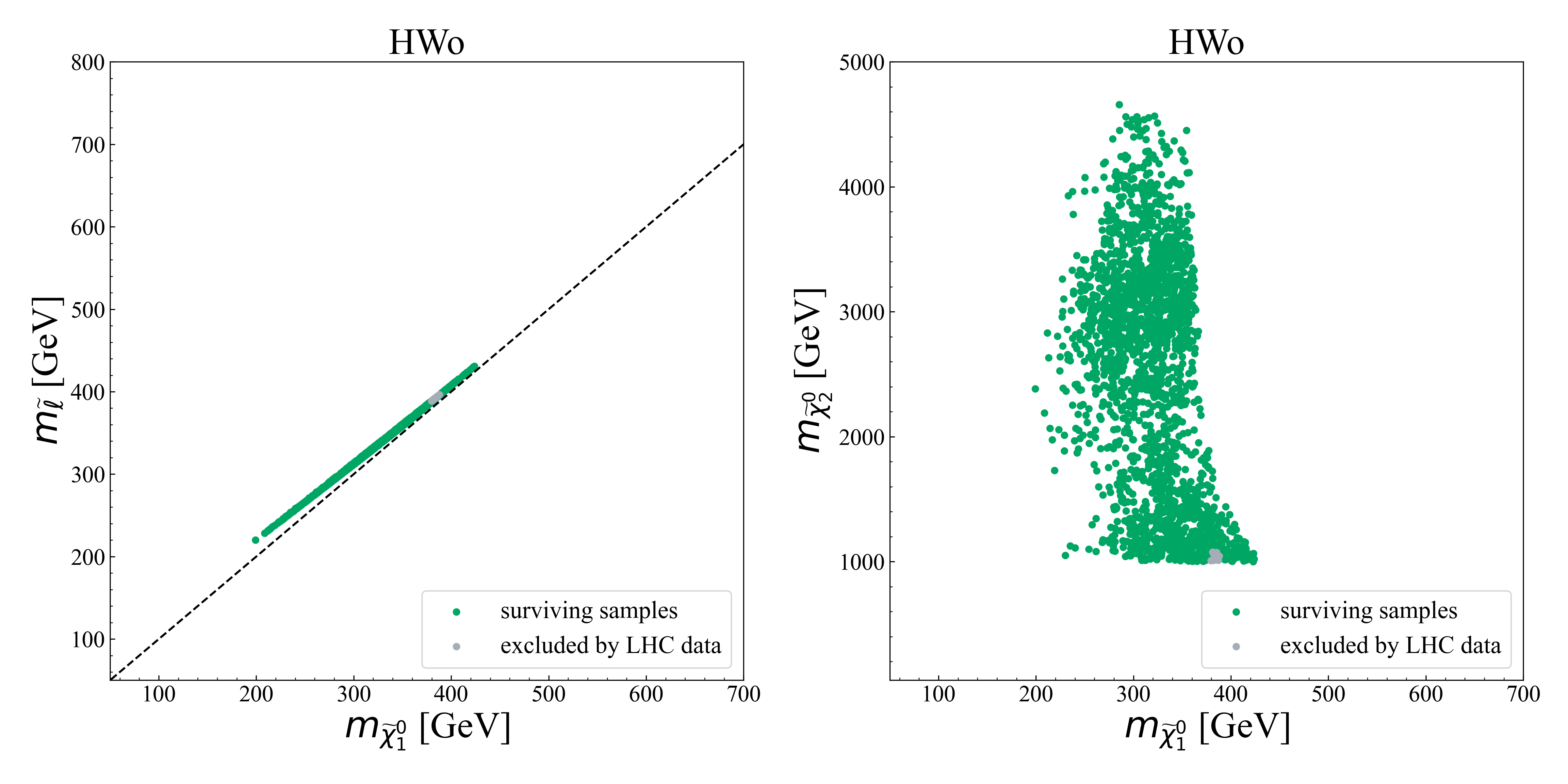}
\vspace{ 0.0cm} 
\caption{
Scatter plots of the samples that survive constraints (1)–(5) and the current direct detection limits, shown in the planes of $m_{\tilde{\chi}1^0}$ versus $m{\tilde{\chi}2^0}$ (left) and $m{\tilde{\ell}}$ (right). The upper and lower panels correspond to the light-wino (LWo) and heavy-wino (HWo) scenarios, respectively.
}
\label{fig:smodels} 
\end{figure*}

In Fig.~\ref{fig:smodels}, we show the samples that survive constraints (1)–(6), and examine their compatibility with current LHC limits using \texttt{SModelS-3.00}~\cite{Altakach:2024jwk}. We summarize the implications for the two scenarios below:
\begin{itemize}
  \item \textbf{For the LWo scenario}:
  \begin{itemize}
    \item The bino-like $\tilde{\chi}_1^0$ and wino-like $\tilde{\chi}_2^0$ have nearly degenerate masses, while slepton masses can either be close to or significantly heavier than that of $\tilde{\chi}_1^0$.
    \item Parameter points with $m_{\tilde{\chi}_1^0} \lesssim 265~\text{GeV}$ are excluded by current LHC SUSY searches. The allowed range for the $\tilde{\chi}_1^0$ mass extends up to around $580~\text{GeV}$.
    \item Slepton masses are further constrained to $285~\text{GeV} \lesssim m_{\tilde{\ell}} \lesssim 630~\text{GeV}$, while the wino-like $\tilde{\chi}_2^0$ must lie in the range $290~\text{GeV} \lesssim m_{\tilde{\chi}_2^0} \lesssim 600~\text{GeV}$.
    \item The strongest exclusion bounds arise from the ATLAS search for chargino–neutralino pair production in trilepton final states with missing transverse momentum (ATLAS-SUSY-2019-09~\cite{ATLAS:2021moa}).
    \item Subdominant exclusions come from the ATLAS analysis targeting compressed electroweakino spectra (ATLAS-SUSY-2018-16~\cite{ATLAS:2019lng}) and from the CMS electroweak chargino–neutralino search (CMS-SUS-17-004~\cite{CMS:2018szt}).
  \end{itemize}

  \item \textbf{For the HWo scenario}:
  \begin{itemize}
    \item In this case, the bino-like $\tilde{\chi}_1^0$ is nearly degenerate with the sleptons, while the wino-like $\tilde{\chi}_2^0$ is significantly heavier.
    \item The mass splitting between $\tilde{\chi}_1^0$ and sleptons is smaller than in the LWo case, leading to a compressed spectrum.
    \item Current LHC limits impose only weak constraints on this scenario.
    \item The most relevant exclusion comes from the ATLAS electroweakino search with two- or three-lepton final states (ATLAS-SUSY-2016-24~\cite{ATLAS:2018ojr}).
  \end{itemize}
\end{itemize}

\section{\label{sec:LHC}Probing weak gauginos at the HL-LHC and HE-LHC}

For parameter points not excluded by \texttt{SModelS}, we evaluate their discovery potential at future colliders: the High-Luminosity LHC (HL-LHC) with a center-of-mass energy $\sqrt{s} = 14\,\text{TeV}$ and integrated luminosity of $L = 3000\,\text{fb}^{-1}$, and the High-Energy LHC (HE-LHC) operating at $\sqrt{s} = 27\,\text{TeV}$ with $L = 15\,\text{ab}^{-1}$.
For parameter points not excluded by \texttt{SModelS}, we evaluate their discovery potential at future colliders: the High-Luminosity LHC (HL-LHC) with a center-of-mass energy $\sqrt{s} = 14\,\text{TeV}$ and integrated luminosity of $L = 3000\,\text{fb}^{-1}$, and the High-Energy LHC (HE-LHC) operating at $\sqrt{s} = 27\,\text{TeV}$ with $L = 15\,\text{ab}^{-1}$.

The dominant SM backgrounds arise from diboson production ($WW$, $WZ$, $ZZ$), $Z$+jets ($Zj$), and top-quark associated processes ($t\bar{t}$ and $Wt$). Signal and background events are generated using the \texttt{MadGraph5\_aMC@NLO}~\cite{Alwall:2014hca}, interfaced with \texttt{Pythia}\cite{Sjostrand:2014zea} for parton showering and hadronization. Detector effects are modeled using \texttt{Delphes}\cite{deFavereau:2013fsa}, and event-level analyses are performed with \texttt{CheckMATE2}~\cite{Drees:2013wra,Kim:2015wza,Dercks:2016npn}.
The statistical significance is estimated using the following expression:
\begin{equation}
\mathcal{Z} = \frac{S}{\sqrt{S + B + (\beta B)^2}},,
\end{equation}
where $S$ and $B$ are the expected numbers of signal and background events, respectively, and $\beta$ denotes the systematic uncertainty. Throughout this analysis, we adopt a conservative estimate of $\beta = 10\%$.

\begin{figure*}[htbp] 
    \centering 
    \includegraphics[width=.9\linewidth]{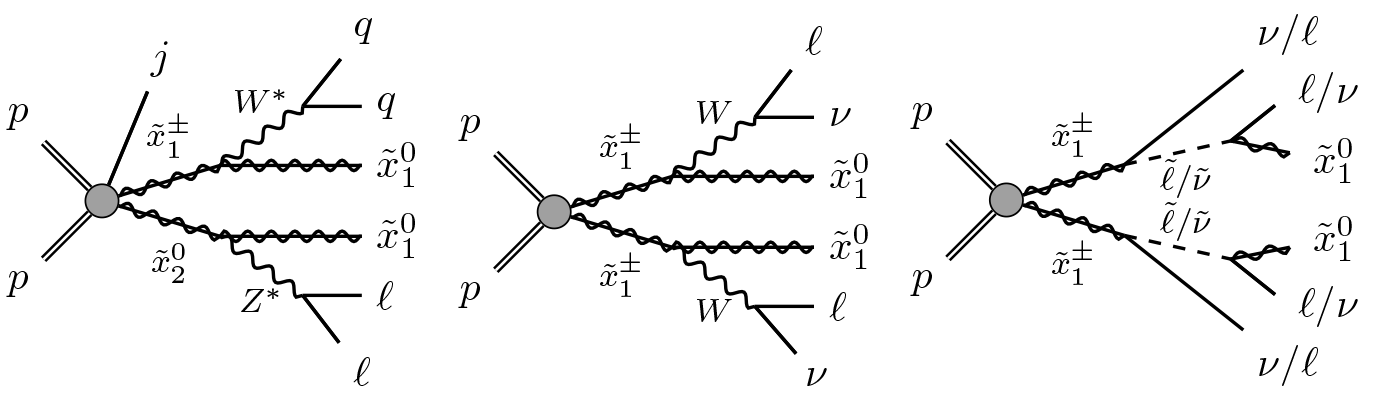}
\vspace{ 0.0cm} 
\caption{
Representative Feynman diagrams for the production processes at the LHC: 
$pp \to j\, \tilde{\chi}_2^0\, \tilde{\chi}_1^\pm$ (left panel), relevant to the LWo scenario, and 
$pp \to \tilde{\chi}_1^\pm\, \tilde{\chi}_1^\mp$ (middle and right panels), relevant to the HWo scenario.
}
\label{fig:fenman-diagrams} 
\end{figure*}

\begin{figure*}[htbp] 
    \centering 
    \includegraphics[width=.9\linewidth]{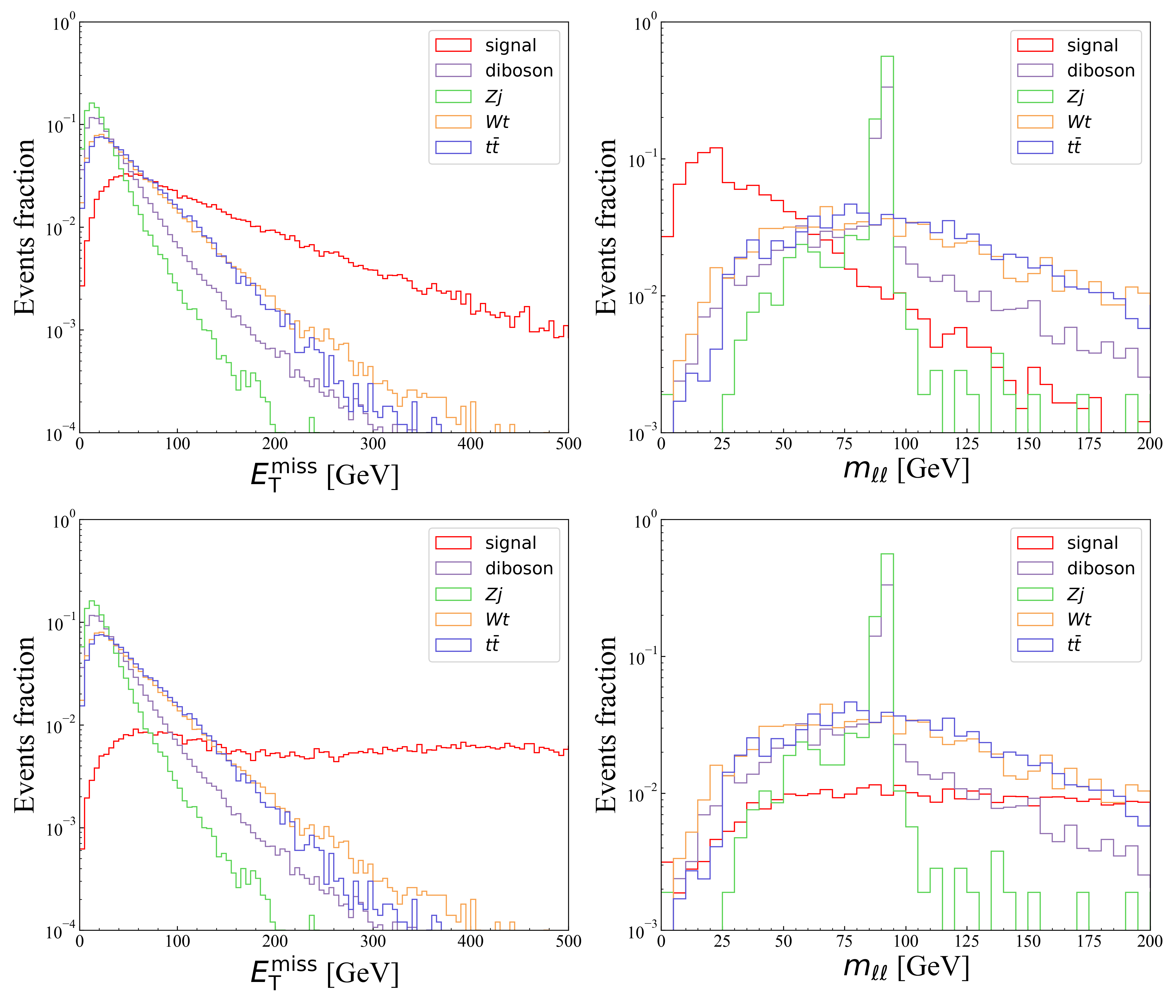}
\vspace{ 0.0cm} 
    \caption{
Normalized distributions of $E_\text{T}^{\text{miss}}$ and $m_{\ell\ell}$ for the signal and background events at the 27~TeV HE-LHC. 
The upper panel corresponds to the \textup{LWo} scenario, with a benchmark point defined by $m_{\tilde{\chi}_1^0} = 375.6~\text{GeV}$, $m_{\tilde{\chi}_2^0} = 401.8~\text{GeV}$, and $\tan\beta = 49.1$. 
The lower panel corresponds to the \textup{HWo} scenario, with $m_{\tilde{\chi}_1^0} = 197.9~\text{GeV}$, $m_{\tilde{\chi}_2^0} = 1072.8~\text{GeV}$, and $\tan\beta = 7.4$.
}
\label{fig:sig-bkg} 
\end{figure*}

In Fig.~\ref{fig:sig-bkg}, the normalized distributions of $E_\text{T}^{\text{miss}}$ and $m_{\ell\ell}$ are shown for both signal and background events. Compared to the backgrounds, both signal scenarios exhibit enhanced event rates in the high-$E_\text{T}^{\text{miss}}$ region. Applying cuts based on this feature can effectively suppress SM backgrounds, particularly the $Zj$ background. Additionally, the LWo signal displays a concentration of dilepton invariant mass events at lower $m_{\ell\ell}$ values, whereas the HWo signal tends to favor larger invariant masses.
Notably, the diboson and $Zj$ backgrounds exhibit a pronounced peak near $m_{\ell\ell} \approx 91$~GeV. To suppress these backgrounds, selection cuts should avoid this specific mass region. Based on these distinct kinematic features, different event selection strategies are applied for the two scenarios.

The event selection criteria for the LWo scenario are defined as follows:
\begin{itemize}
\item \textbf{Cut-1:} Require missing transverse energy $E_\text{T}^{\text{miss}} > 200$~GeV.

\item \textbf{Cut-2:} Select events with two opposite-sign (OS), same-flavor (SF) leptons (e.g., $e^+$ and $e^-$). The leading lepton must satisfy $p_\mathrm{T}^{\ell_1} > 5$~GeV.

\item \textbf{Cut-3:} To suppress the $t\bar{t}$ background, events with any identified $b$-jets are vetoed. Additionally, require at least one jet with leading transverse momentum $p_\mathrm{T}^{j_1} > 100$~GeV. The azimuthal angle difference between the leading jet and missing momentum must satisfy $\Delta\phi(j_1, \mathbf{p}_\text{T}^{\text{miss}}) > 2$, and for all jets, $\Delta\phi(j, \mathbf{p}_\text{T}^{\text{miss}}) > 0.4$.

\item \textbf{Cut-4:} To further reduce backgrounds, impose the following requirements:
\begin{itemize}
    \item Veto events with reconstructed di-tau mass $m_{\tau\tau} \in [0, 160)$~GeV.
    \item Require dilepton invariant mass $1~\text{GeV} < m_{\ell\ell} < 60~\text{GeV}$, excluding the $J/\psi$ resonance region by $m_{\ell\ell} \notin (3, 3.2)$~GeV.
\end{itemize}

\item \textbf{Cut-5:} Apply additional kinematic criteria:
\begin{itemize}
    \item Define the scalar sum of lepton transverse momenta as $H_\text{T}^{\text{lep}} = p_\text{T}^{\ell_1} + p_\text{T}^{\ell_2}$, and require $E_\text{T}^{\text{miss}} / H_\text{T}^{\text{lep}} > \max(5,\; 15 - 2 m_{\ell\ell})$.
    \item Require angular separation between leptons $0.05 < \Delta R_{\ell\ell} < 2$.
    \item The transverse mass of the leading lepton must satisfy $m_\text{T}^{\ell_1} < 70$~GeV.
\end{itemize}
\end{itemize}

Table~\ref{tab:cutflow1} summarizes the cutflow results for the LWo scenario, detailing the impact of each successive selection. As shown in the table, Cut-2 substantially suppresses most SM backgrounds, particularly the $Zj$ process, consistent with findings in Refs.~\cite{ATLAS:2019lff, ATLAS:2017vat}. The implementation of Cut-3 and Cut-4 further reduces all backgrounds while preserving a high signal efficiency. Finally, Cut-5 proves especially effective at suppressing residual backgrounds, although a small fraction of $Zj$ events still survive.

\begin{table}[h]
\centering
\begin{tabular}{
  >{\centering\arraybackslash}p{2cm} 
  >{\centering\arraybackslash}p{2cm} 
  c c c c 
}
\toprule
Cuts & LWo & ${t}\bar{{t}}$ & diboson & $Zj$ & $Wt$   \\ 
\midrule
Cut-1 & 68.45 & 38025.12 & 1738.03 & 49369.14 & 1654.35 \\  
Cut-2 & 5.76 & 828.71 & 38.06 & 90.07 & 47.76 \\  
Cut-3 & 3.75 & 64.83 & 15.89 & 31.94 & 6.43 \\  
Cut-4 & 3.31 & 22.81 & 4.14 & 4.26 & 2.09 \\  
Cut-5 & 0.95 & 2.15 & 0.41 & 1.19 & 0.21 \\  
\bottomrule
\end{tabular}
\caption{Cut flow of the cross sections (in fb) for the signal and SM backgrounds at the 27 TeV HE-LHC, based on the LWo benchmark point with $m_{\tilde{\chi}1^0} = 375.6\,\text{GeV}$, $m{\tilde{\chi}_2^0} = 401.8\,\text{GeV}$, and $\tan\beta = 49.1$.}
\label{tab:cutflow1}
\end{table}

The event selection criteria for the HWo scenario are defined as follows:
\begin{itemize}
\item \textbf{Cut-a:} Exactly two opposite-sign, same-flavor leptons ($N(\ell) = 2$) are required, each with transverse momentum $p_\mathrm{T}^{\ell_{1,2}} > 25~\text{GeV}$. The dilepton invariant mass must satisfy $m_{\ell\ell} > 25~\text{GeV}$. In addition, we require no reconstructed $b$-jets and fewer than one non-$b$-tagged jet ($N_{\text{non-b-jet}} < 1$) to suppress single-top ($Wt$) and $t\bar{t}$ backgrounds.

\item \textbf{Cut-b:} To eliminate low-mass resonances and suppress $Z$-boson-related backgrounds ($WZ$, $ZZ$, $Zj$), the dilepton invariant mass must satisfy $m_{\ell\ell} > 121.2~\text{GeV}$, thereby avoiding the $Z$-mass peak.

\item \textbf{Cut-c:} The missing transverse energy is required to exceed $E_\text{T}^{\text{miss}} > 110~\text{GeV}$.

\item \textbf{Cut-d:} The missing transverse energy significance, as defined in Ref.~\cite{ATLAS:2018uid}, must satisfy $E_\text{T}^{\text{miss}}~\text{significance} > 10$. Additionally, the stransverse mass is required to be $m_{\text{T2}} > 100~\text{GeV}$, where $m_{\text{T2}}$ is defined in Refs.~\cite{Lester:1999tx,Barr:2003rg,Lester:2007fq}.

\end{itemize}

Table~\ref{tab:cutflow2} summarizes the evolution of background and signal cross sections after each selection step in the HWo analysis. The dominant $Zj$ background is substantially reduced by the dilepton invariant mass cut in Cut-b, and is rendered negligible after the missing energy requirement in Cut-c. The final kinematic selection (Cut-d) efficiently eliminates the residual background while maintaining high signal sensitivity.

\begin{table}[h]
\centering
\begin{tabular}{
  >{\centering\arraybackslash}p{2cm} 
  >{\centering\arraybackslash}p{2cm} 
  c c c c 
}
\toprule
Cuts & HWo & ${t}\bar{{t}}$ & diboson & $Zj$ & $Wt$   \\ 
\midrule
Cut-a & 172.09 & 297932 & 812800 & 75903863 & 34182.18 \\  
Cut-b & 150.97 & 142463 & 157933 & 1371.44 & 16796.26 \\  
Cut-c & 138.85 & 20817.40 & 8557.80 & 0 & 2119.84 \\  
Cut-d & 106.10 & 238.35 & 15.47 & 0 & 18.77 \\  
\bottomrule
\end{tabular}
\caption{Cut flow of the cross sections (in ab) for the signal and SM backgrounds at the 27~TeV HE-LHC, based on the HWo benchmark point with $m_{\tilde{\chi}_1^0} = 197.9~\text{GeV}$, $m_{\tilde{\chi}_2^0} = 1072.8~\text{GeV}$, and $\tan\beta = 7.4$.}
\label{tab:cutflow2}
\end{table}

\begin{figure*}[htbp] 
    \centering 
    \includegraphics[width=.9\linewidth]{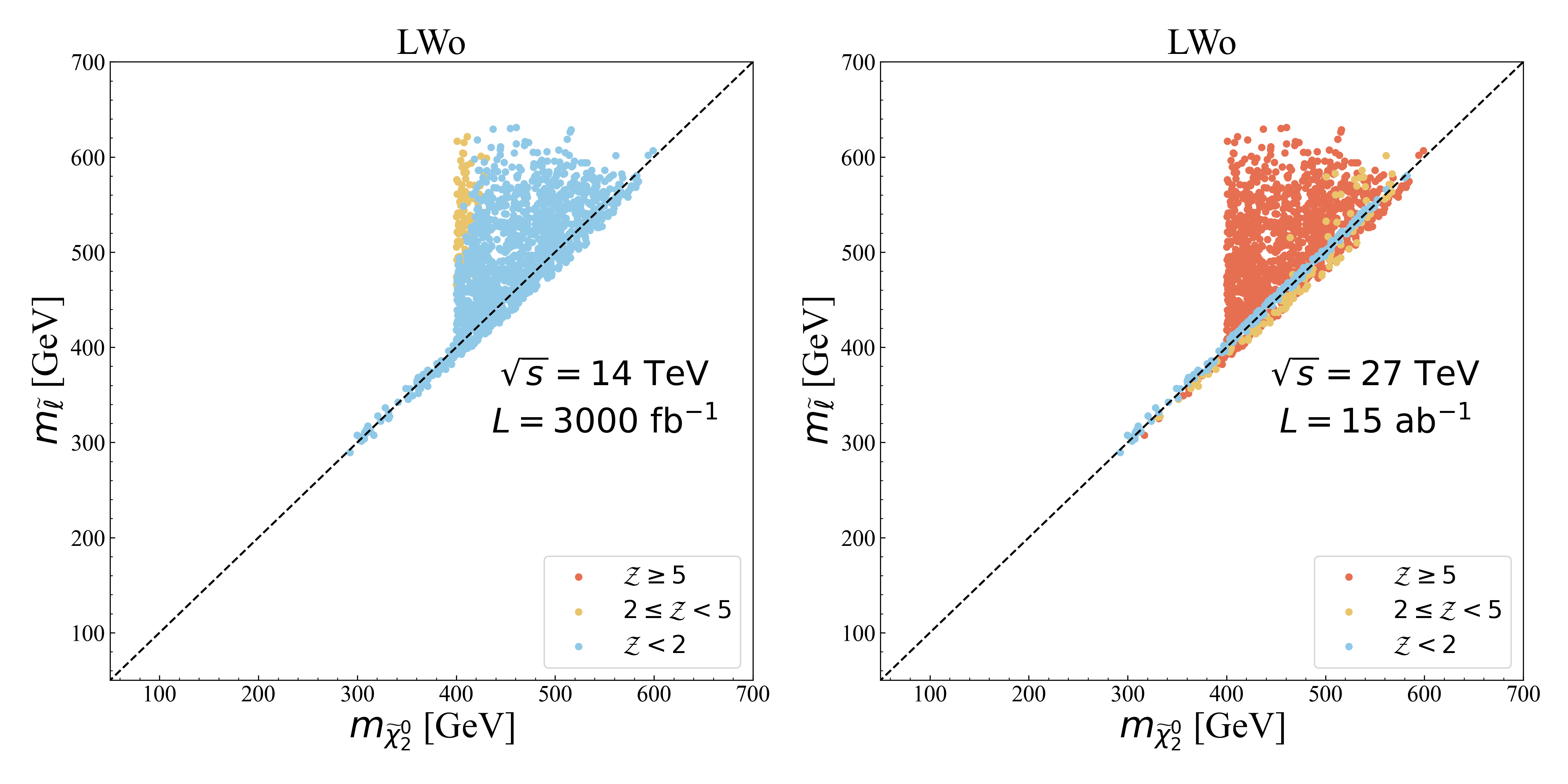}\hspace{.1\linewidth}
    \includegraphics[width=.9\linewidth]{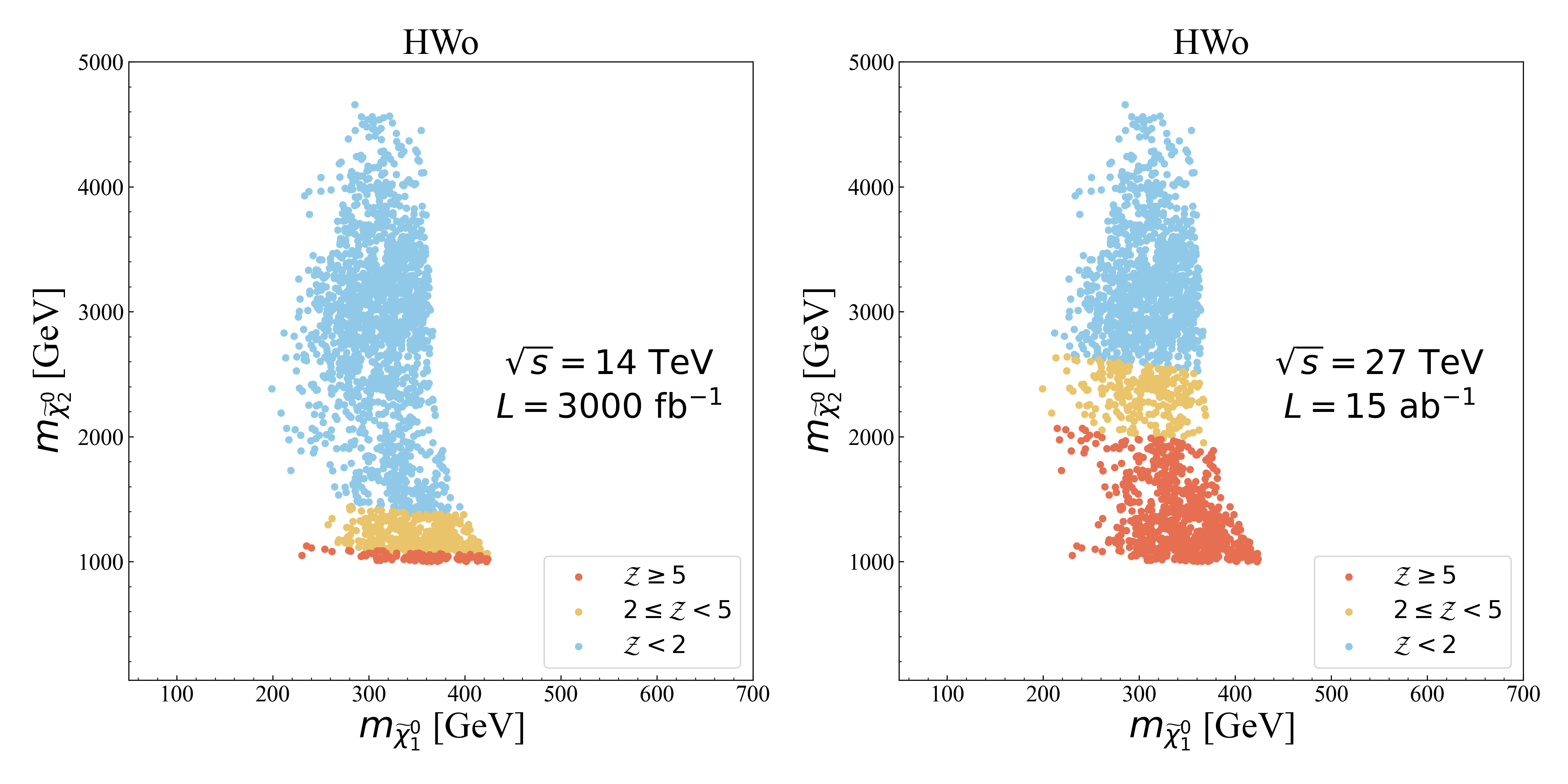}
\vspace{ 0.0cm} 
\caption{Significance of the surviving parameter points from Fig.~\ref{fig:smodels} at the HL-LHC and HE-LHC. The upper and lower panels correspond to the LWo and HWo scenarios, respectively.}
\label{fig:significance} 
\end{figure*}

In Fig.~\ref{fig:significance}, we present the projected discovery significance of the LWo and HWo scenarios at the HL-LHC and HE-LHC. The main features are summarized as follows:
\begin{itemize}
    \item \textbf{HL-LHC sensitivity:}
    \begin{itemize}
        \item For the \textup{LWo} scenario, only a small fraction of parameter points are expected to be probed at the HL-LHC.
        \item For the \textup{HWo} scenario, most points remain viable. However, regions with relatively small $\tilde{\chi}_2^0$ masses may be within the sensitivity reach of the HL-LHC.
    \end{itemize}

    \item \textbf{HE-LHC sensitivity:}
    \begin{itemize}
        \item In the \textup{LWo} case, a significant portion of parameter space with large mass splitting between the wino-like $\tilde{\chi}_2^0$ and sleptons can be probed. The surviving points tend to lie in the mass-compressed region between $\tilde{\chi}_2^0$ and $\tilde{\ell}$.
        \item For the \textup{HWo} scenario, the viable parameter space is further reduced. In particular, regions where the wino-like $\tilde{\chi}_2^0$ mass lies below approximately 2~TeV are expected to be covered by the HE-LHC.
    \end{itemize}
\end{itemize}

\noindent
In addition to conventional cut-based analyses, applying machine learning techniques~\cite{Abdughani:2019wuv, Caron:2016hib, Guo:2023nfu, Ren:2017ymm, Abdughani:2020xfo} could further improve the sensitivity to supersymmetric particles, potentially refining the reach estimates presented in this study.

\section{\label{sec:conclusions}Conclusions}

We have studied the LHC discovery potential for weak gauginos in the MSSM, motivated by the persistent $5.2\sigma$ discrepancy in the muon anomalous magnetic moment and the latest constraints from dark matter and collider experiments. By scanning the parameter space under bounds from Higgs measurements, relic abundance, direct detection, and LHC simplified model results, we identified two viable scenarios: one with a light wino (LWo), and one with a heavy wino (HWo), both involving a bino-like neutralino as the dark matter candidate.

In both scenarios, the observed dark matter abundance is obtained via co-annihilation, either with winos or sleptons, while maintaining SUSY contributions sufficient to explain the muon $g-2$ anomaly. The LWo case permits a broader mass range, with the LSP and sleptons extending up to 580 GeV and 650 GeV, respectively. In contrast, the HWo scenario is more constrained, with upper bounds near 430 GeV. Both scenarios exhibit compressed spectra, posing challenges for collider searches.
Cut-based analyses at the HL-LHC indicate limited sensitivity, whereas the HE-LHC with $\sqrt{s} = 27$~TeV and $L = 15$~ab$^{-1}$ can significantly improve coverage, probing most of the viable parameter space in both scenarios. The distinct kinematic characteristics of LWo and HWo suggest that tailored strategies are essential for optimizing discovery potential.

Overall, our results indicate that weak gauginos remain viable supersymmetric solutions to the muon $g-2$ anomaly. While the HE-LHC provides promising prospects for probing these scenarios, further refinements, such as dedicated searches targeting compressed spectra or the use of machine learning techniques, could play an important role in improving sensitivity.




\acknowledgments
This work was supported by the National Natural Science Foundation of China under Grant No. 12275066 and by the startup research funds of Henan University. 




\bibliographystyle{JHEP}
\bibliography{biblio.bib}




\end{document}